# An Optical Watermarking Solution for Color Personal Identification Pictures

Tan Yi-zhou*[a], Liu Hai-bo[b], Huang Shui-hua[b], Sheng Ben-jian[b], Pan Zhong-ming[a]

[a] College of Mechatronics Engineering and Automation; [b]College of Science,
National University of Defense Technology, Changsha, Hunan, China, 410073

## ABSTRACT

This paper presents a new approach for embedding authentication information into image on printed materials based on optical projection technique. Our experimental setup consists of two parts, one is a common camera, and the other is a LCD projector, which project a pattern on personnel's body (especially on the face). The pattern, generated by a computer, act as the illumination light source with sinusoidal distribution and it is also the watermark signal. For a color image, the watermark is embedded into the blue channel. While we take pictures (256 ×256 and 512×512, 567×390 pixels, respectively), an invisible mark is embedded directly into magnitude coefficients of Discrete Fourier transform (DFT) at exposure moment. Both optical and digital correlation is suitable for detection of this type of watermark. The decoded watermark is a set of concentric circles or sectors in the DFT domain (middle frequencies region) which is robust to photographing, printing and scanning. The unlawful people modify or replace the original photograph, and make fake passport (drivers' license and so on). Experiments show, it is difficult to forge certificates in which a watermark was embedded by our projector-camera combination based on analogue watermark method rather than classical digital method.

## 1. INTRODUCTION

A very common mode of unlawful actions is via unauthorized distribution of hard-copy documents. In recent years, with some image editing tools (for example Photoshop), it is not hard for professionals to replace images in stolen passports with new images for unlawful actions. In order to avoid the weaknesses of the current passport or other identification documents, some watermarking methods for print-scan resilient data hiding in the photograph were proposed [1][2][3]. For example, Anthony T. S. Ho and Feng Shu establish a linkage between the owner's photo and his/her personal information by superimposing the ID numbers into the original photograph [1]. Emir Ganic and Scott D. Dexter generalized a circular watermarking idea to embedding multiple watermarks that are not visible in the DFT domain [2].

Most existing watermark schema in document authentication systems depends on the digital image processing programs in the complete process, from embedding information to watermarking detection. In this paper, a document authentication scheme based on hybrid analogue/digital watermark is presented. At the camera exposure moment we use projector-camera combination (analogue technique) to generate an invisible mark in the photograph's Discrete Fourier transform(DFT) spectrum, and in the watermark detection process we use digital image processing technique, which is the same as reported in the literatures [3][4][5]. The experiment results shown that the embedded information on printed materials is sufficiently robust that it can be reliably extracted after image distortion, rotation and scaling.

## 2. DOCUMENT AUTHENTICATION SCHEME

### 2.1 Watermark embedding scheme based on digital image processing technique

Solachidis and Pitas embed a 2D circularly symmetric sequence in a ring covering the middle frequencies,

where the ring is separated in multiple sectors [4]. Licks and Jordan use a watermark in the form of a circle with a radius that corresponds to higher frequencies of the image [5].

In the literature [4] and [5] an circularly symmetric watermark is embedded in the DFT domain by modifying the highest energy DCT coefficients $M(u,v)$ using the additive Formula

$$MW(u,v) = M(u,v) + aW(u,v), \tag{1}$$

where $M$ denotes the magnitude of DFT coefficients of the image (for example, Passport photo), and $W$ is the circular watermark. $a$ is the watermark strength also directly influencing watermark visibility.

The watermark detection scheme is as following:

- Subtract the original image from the watermarked image (Passport photo), and extract the watermark sequence η' (may be corrupted due to image distortion)

- Correlate η with η' using the Formula

$$\text{sim}(\eta, \eta') = \frac{\eta \cdot \eta'}{\sqrt{\eta' \cdot \eta'}} \tag{2}$$

**sim**(η, η') is called similarity,

**sim**(η, η') > *Th* => watermark is present,

**sim**(η, η') < *Th* => watermark is not present,

Where *Th* denotes the threshold, and η = original watermark sequence.

## 2.2 Hybrid analogue/digital watermarking scheme

In Figure 1，the main equipments of watermarking unit are scanner, printer and computer in literature [1] and [4]. The watermark is generated by an embedding algorithm mentioned in §2.1.We have the ability to embedding a watermark into DCT coefficients of the photo, only to have original photograph (analogue signal) converted to digital signal with the scanner just before watermarked image is generated with the computer.

In this paper we choose the same watermarking principle as §2.1 with different equipments, which is shown in Fig.2. Our hybrid analogue/digital watermarking setup consists of two equipments, one is a camera (not specialized), and the other is a liquid crystal display (LCD) projector, which project a pattern on the personnel's body, especially on the face. Taking a picture and embedding watermark take place at the same time.

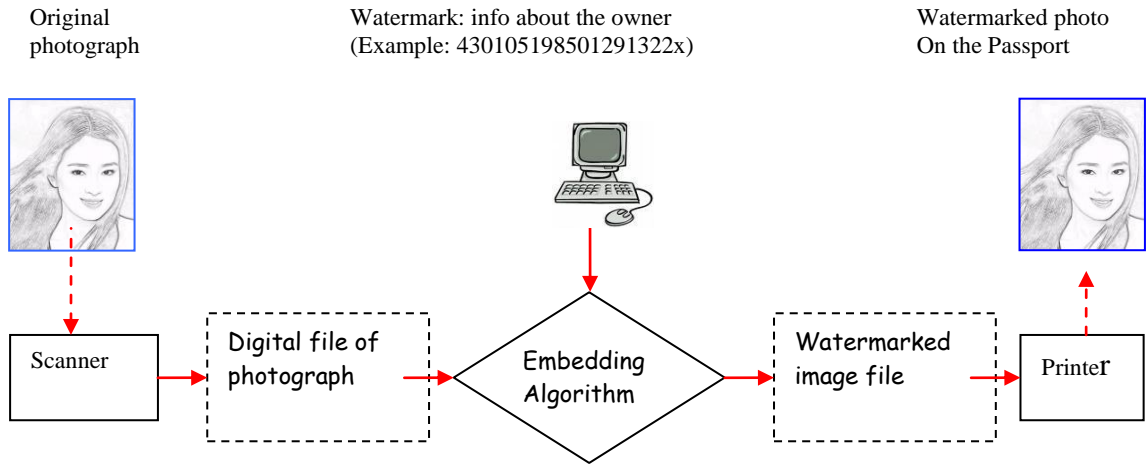

Figure 1 Watermark embedding scheme based on classical digital image processing technique

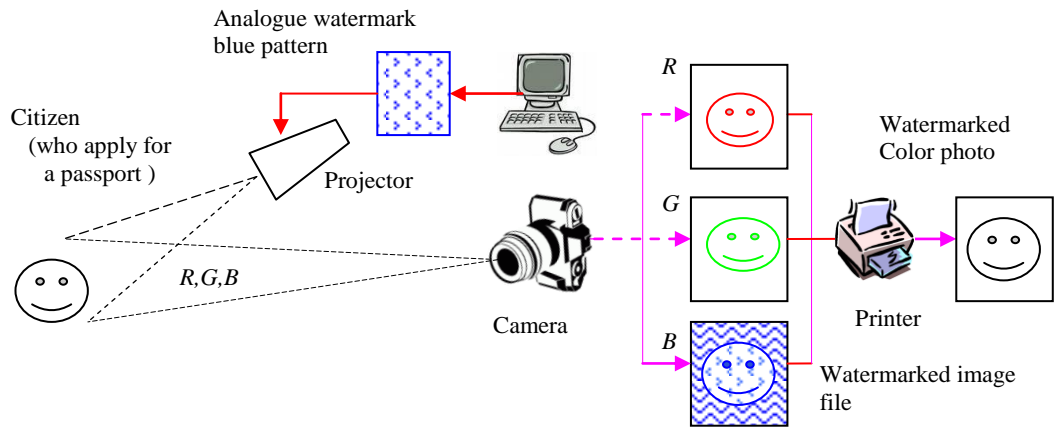

Figure 2  Watermark embedding scheme based on analogue technique using the projector-camera combination

# 3. DOCUMENT AUTHENTICATION EXPEREMENTS

## 3.1 Photogrammentric system

Unlike the common indoor photogrammentric system which consists of a camera and a set of illumination light sources, our photogrammentric system includes the camera, the LCD projector and computer, as shown in Figure 3. The LCD projector projects white light to illuminate the personnel's body and the screen (or background). In Figure4, the red and green channels are homogeneous beams; the blue channel carries a watermark pattern.

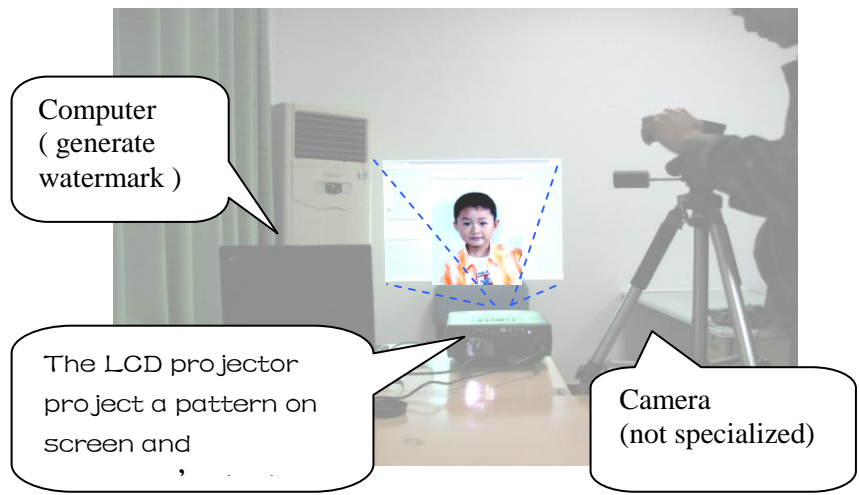

Figure 3 Watermark embedding setup (photogrammentric system with projector-camera combination)

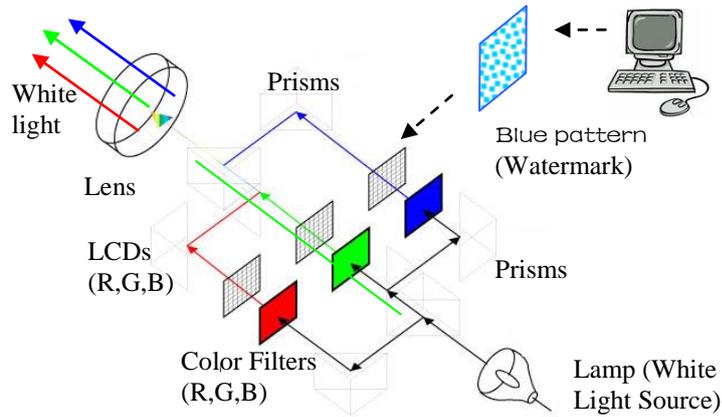

Figure 4  Red ,Green beams and Blue pattern are mixed into white light by the lens in LCD projector

### 3.2  Experiment results

Before photographing, we generate a blue pattern by Discrete Fourier Transform Algorithm. The Figure 5(b) shows the coding pattern in the Blue band of the projector, and the Figure 5 (d) shows its DFT magnitude spectrum. Using a set of sinusoids figures with different frequencies and orientations to denote the owner's ID numbers the watermark has been embedded in a set of rings that covers the middle frequencies region.

The photo with common illumination lamp and the photo with our projector-camera combination are shown in Fig. 5 (a) and (c), respectively. As can be seen, the watermark (contain 16-24 binary numbers in 567*390 pixels) is invisible in the color cover image. Although the personnel (especially his face) has been coded by the blue illumination light, the actual illumination effect of three color beams emitted from the projector Fig 4 is equivalent to white light emitted from a common lamp.

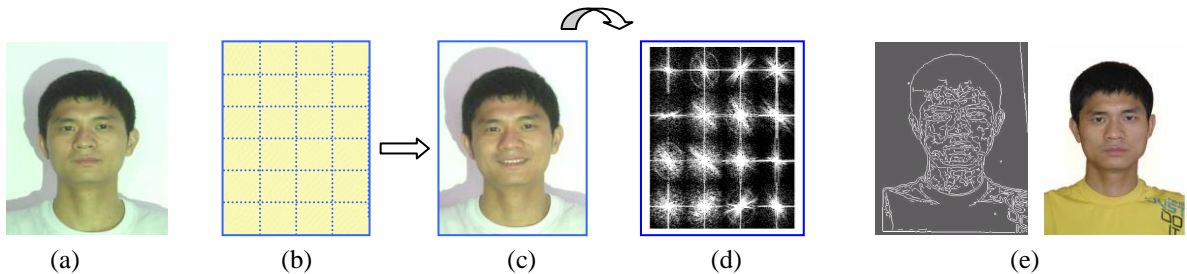

(a)                (b)                (c)                (d)                (e)

Figure 5 Experiment results of photogrammentric system with projector-camera combination

(a) Common photo, (b) The coding pattern of the Blue band of the projector,  (c) The watermarked photo with the LCD projector in Fig. 4 as the illumination source,  (d) Discrete Fourier transform spectrum of the watermark pattern  (e) Clear up the watermark on the outside of body's image based on Canny Operator

Some countries have special requirements for the certificate photo size and color of background. Photographs for passport must have a plain, light-coloured background (e.g. white, cream or pale blue). The light-coloured background is unsuitable for embedding watermark. After taking a picture with photogrammentric system (Fig.3) post processor program is need to clear up the watermark on the outside of body's image.  As shown in Fig5 (e), geometric contour of body was extracted based on Canny Operator. The computer program uses the contour as the edge-restraint condition to ensure that watermark not present at background.

### 3.3  Robustness of the watermarking method based on the projector-camera combination

In the watermark embedding process, we used a common color ink-jet printer/scanner (HP PSC 1218) to print the digital image on security document (the owner's passport or identification card). In the watermark decoding process the printed photo is scanned to obtain a digital image file for watermark detection. We have tested on several photographing conditions showing that the embedded signal in Fourier amplitude spectrum is sufficient to be detected by desk top scanners (or camera with near-field lens).

It is well known that the rescaling operation has almost no effect on the DFT coefficients, and that after rotating an image in the spatial domain, the locations of the coefficient values will have the same rotation in the DFT domain. Watermark extraction experimental results show the effectiveness of image scaling, rotation (if the rotation and trapezoidal distortion had been corrected by digital image processing programs).

.Because the watermark in DFT `spectrum` was spread over the full area, information loss in its any local part have insignificant influence on detecting watermark so that the scratched or smeared photos( in Fig.6) can be smoothly decode ..

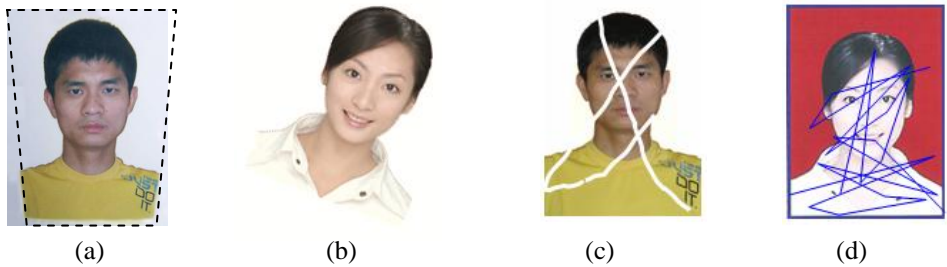

(a)                    (b)                    (c)                    (d)

Figure 6 Robustness of the projector-camera combination watermarking method

(a) Rotation photo, (b) Trapezoidal distortion, (c) The scratched photo, (d) Smeared with ink.

Fourier spectrum image noise with relation to print-and-scan is introduced. Because noise appeared in low frequencies region caused by printer and scanner, its contrast of the watermark rings blurred (as shown in Fig. 7(lift)). This phenomenon indicate that the modulate strength of the DFT magnitude coefficients near the strength threshold.

In our experiments, it is important to select higher modulation strength than the threshold for tradeoff between the robustness and image quality (includes decoded watermark image contrast).

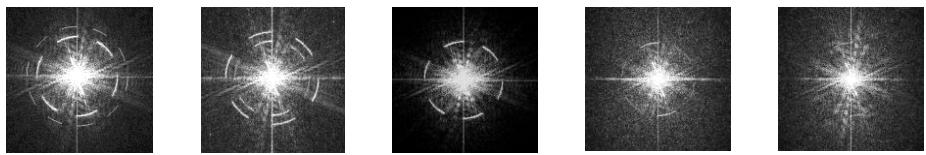

Figure 7 In Fourier spectrum image the Signal-to-Noise Ratio of the decoded watermark is increasing from the lift to right with relation to print-and-scan processes and photographing processes.

## 4. CONCLUSION

In this paper, we propose a watermarking scheme that uses LCD projector-camera combination but not computer to embed digital watermark into photos at photographing processes. The experimental results shown that it is possible to embed a circularly symmetric watermark in the DFT domain using hybrid analogue-digital data hiding technique. And the color photos have acceptable image values for owner identification and robustness against the printing and scanning. The embedded signals in photo can be detected by optical or digital correlation (inverse transformation) at relatively low Signal-to-Noise Ratio in the Discrete Fourier transform spectrum.

Because both the personnel's image and background's image suffered different influences of illumination condition and the modified strength of the DFT coefficients, the reproducibility of analogue watermark is lower than that of digital watermark. One advantage of the low reproducibility is that the degree of difficulty for forging and altering photos on document and certificates tends to increase, if we use "pure" optical technique to embed analogue watermark into the photograph at its exposure moment.

# 5. ACKNOWLEDGEMENTS

This work has been supported by the Innovation Experiment & Inquiry Learning Program for University Students of Provincial Education Department of Hunan, China. No. 4-2008.

## REFERENCE LINKING